ARTICLE

# 2D MXene Electrochemical Transistors

Jyoti Shakya,[a†] Min-A Kang,[ab†], Jian Li,[a] Armin VahidMohammadi,[c] Weiqian Tian,[ad]* Erica Zeglio,[ef]* Mahiar Max Hamedi[a]*

In the past two decades another transistor based on conducting polymers, called the organic electrochemical transistor (ECT) was shown and largely studied. The main difference between organic ECTs and FETs is the mode and extent of channel doping: while in FETs the channel only has surface doping through dipoles, the mixed ionic-electronic conductivity of the channel material in Organic ECTs enables bulk electrochemical doping. As a result, the organic ECT maximizes conductance modulation at the expense of speed. Until now ECTs have been based on conducting polymers, but here we show that MXenes, a class of 2D materials beyond graphene, have mixed ionic-electronic properties that enable the realization of electrochemical transistors (ECTs). We show that the formulas for organic ECTs can be applied to these 2D ECTs and used to extract parameters like mobility. These MXene ECTs have high transconductance values but low on-off ratios. We further show that conductance switching data measured using ECT, in combination with other in-situ ex-situ electrochemical measurements, is a powerful tool for correlating the change in conductance to that of redox state: to our knowledge, this is the first report of this important correlation for MXene films. Many future possibilities exist for MXenes ECTs, and we think other 2D materials with bandgaps can also form ECTs with single or heterostructured 2D materials. 2D ECTs can draw great inspiration and theoretical tools from the field of organic ECTs and have the potential to considerably extend the capabilities of transistors beyond that of conducting polymer ECTs, with added properties such as extreme heat resistance, tolerance for solvents, and higher conductivity for both electrons and ions than conducting polymers.

[a.] Department of Fibre and Polymer Technology, KTH Royal Institute of Technology, Teknikringen 56, 10044 Stockholm, Sweden
[b.] Department of Materials Science and Engineering, Northwestern University, Evanston, IL 60208, USA
[c.] A. J. Drexel Nanomaterials Institute and Department of Materials Science and Engineering, Drexel University, Philadelphia, PA 19104, USA
[d.] School of Materials Science and Engineering, Ocean University of China, Qingdao, Shandong 266100, China
[e.] Division of Nanobiotechnology, Department of Protein Science, Science for Life Laboratory, School of Engineering Sciences in Chemistry, Biotechnology and Health, KTH Royal Institute of Technology and Digital Futures, Solna, Sweden.
[f.] Center for the Advancement of Integrated Medical and Engineering Sciences (AIMES), Karolinska Institutet and KTH Royal Institute of Technology, 171 77, Stockholm, Sweden
† Equal contribution
Electronic Supplementary Information (ESI) available: See DOI: 10.1039/x0xx00000x

# ARTICLE

## 1. Introduction

The discovery of the basic electronic behavior of one-atom-thick graphene in 2004.[1] gave rise to two important directions in materials science: the field of physics in two-dimensional (2D) materials and the realization of a future roadmap of 2D materials beyond graphene. Additionally, one of the most important technical advances in 2D materials has been the discovery that layered crystals can be exfoliated and colloidally stabilized in liquids, enabling their fabrication from polar solvents, such as water.[2] Computational studies have predicted that thousands of materials can exfoliate into 2D materials[3]; a number that is driven up by the contribution of scientific technologies. Among them, transition metal dichalcogenides (TMDs) and transition metal carbides and/or carbonitrides labeled MXenes[4] currently comprise the large classes of 2D materials beyond graphene. TMDs and MXenes introduce many new capabilities not present in graphene, such as tunable bandgaps and conductivity, and high pseudocapcitance.[5, 6]

The electronic behavior of graphene was initially studied using field effect transistors (FETs).[1] FETs constitute the building block of computers and therefore nano-engineered 2D materials have been pursued as an alternative to metal-oxide-semiconductor FETs (MOSFETs) in complementary metal-oxide-semiconductor (CMOS).[7] In FETs, the induced dipole moments at the dielectric layer generated by the application of a gate bias modulate the electrical field and conductivity across the channel.

The FET as a three-terminal device also opened the path towards another form of transistor popularized in the past two decades: the organic electrochemical transistor (OECT).[8]

Similar to electrolyte-gated FETs, OECTs include an electrolyte between the channel and gate. The main difference between OECTs and FETs is the mode and extent of channel doping: while in FETs the channel only has surface doping through dipoles, the mixed ionic-electronic conductivity of the channel material in OECTs enables bulk electrochemical doping.[9] As a result, the OECT maximizes conductance modulation at the expense of speed.[10] Organic mixed ionic/electronic conductors, such as conjugated polymers are the most common OECT materials, with conductance switching generated by modulation of their redox state upon doping.[11] The most dominant OECT material to date is poly(3,4-ethylenedioxythiophene):polystyrene sulfonate (PEDOT:PSS), due to its stability in water and ambient conditions (e.g., temperature and presence of oxygen), large capacitance, and high ionic conductivity. PEDOT:PSS operation hinges on the decrease of conductance through the depletion of mobile holes upon cation injection.[12] While PEDOT:PSS represents the most prominent example of a p-type organic conductor, n-type materials have emerged in the past decade, driven by the need to increase device complexity and sensitivity for biosensors. In contrast to p-type materials, in n-type semiconductors the conductivity is provided by the formation of mobile electrons upon cation injection.[13]

OECTs based on conducting polymers have been applied to several fields: electronic textiles,[14] biology,[15] displays, and neuromorphic computers.[10, 16] However, the demonstration of electrochemical transistors based on inorganic materials (herein referred to as ECTs) has yet to be explored. In this manuscript we will thereby use the term OECTs to refer to organic ECTs, while electrochemical transistors based on 2D materials will be referred to as 2D ECTs.

We recently described that 2D titanium carbide ($Ti_3C_2T_x$) MXene films change their bulk electronic conductivity at low operating voltages (similar to doping in conducting polymers) and used 2D MXene to realize electrochemical random access memories (ECRAMs).[17] Our results showed that MXene films could act as mixed ion-electron conductors for this specific application.

Here, we considerably advance this work by proposing that these MXene devices indeed belong to the larger class of ECT devices. We further evaluate the impact of controlled MXene thin film formation on the redox properties and mixed ionic/electronic conductivity of films by fabricating the ECTs using layer-by-layer (LbL) assembly. We further show that the theories for ECTs can be applied to MXenes and provide, to the best of our knowledge, the first experiments that elucidate the switching mechanism of a 2D material using a combination of spectroelectrochemistry and ECT device characterization.

## 2. Results and discussion

### 2.1 Fabrication of multi-layered 2D MXene and characterization

Ultrathin 2D $Ti_3C_2T_x$ MXene films have been used to fabricate field-effect transistors (FETs).[18] However, to achieve

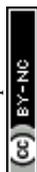

Nanoscale Accepted Manuscript

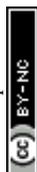

the volumetric capacitance needed for ECTs, we need to go beyond a single layer. This requires implementing scalable methods to form unique multilayered stacks. To obtain the best possible performance in terms of electronic conductivity, these layered films should form well-ordered stacks so that adjacent 2D layers have appropriate geometrical sizes and maximum interfacial contact. Moreover, to provide high capacitance, the chemical functionalities and spacing between the 2D flakes should be tailored for efficient ion transport and ion-electron coupling. We have previously shown that directed layer-by-layer (LbL) self-assembly of 2D materials, in which no polymer but only a small charged molecule is used between the layers, is a viable option for fabricating multi-layered 2D materials with volumetric capacitance up to 583 F cm$^{-3}$.[19] In short, this method relies on the deposition of alternating layers of oppositely charged materials from an aqueous solution at room temperature. A positively charged substrate is immersed in a solution of negatively charged 2D material flakes leading to the formation of a very thin layer, held together by electrostatic interactions. Further immersion in a solution of a positively charged molecule reverses the surface charge enabling the deposition of another layer of 2D material. The process is repeated n times to form n-layered films. In this study, we used an aqueous solution of the positively charged tris(3-aminopropyl)amine (TAPA) at pH 7.5 for the controlled fabrication of 2D $Ti_3C_2T_x$ MXene multilayers. We compared this fabrication method with spin-coated MXene films to better understand the impact of controlled layer formation on the redox properties and mixed ionic/electronic conductivity of the films. For spin-coated films, 5 spin-coating cycles of a MXene dispersion (2 g L$^{-1}$ in water, see experimental section for details) were needed to fully and homogeneously cover the substrate, whereas, for LbL-assembled films, we could achieve full coverage already for 3 MXene/TAPA bi-layers (Figure 1a). Field Emission Scanning Electron Microscope (FESEM) images display representative cross-sectional images for both spin-coated (Figure 1b) and LbL-assembled (Figure 1c) MXene on Si substrates. These images show that films produced via LbL-assembly with n = 20 exhibit higher thickness and layers organization with respect to films prepared by subsequent spin-coating steps (20 in total, see experimental section for details). We previously used FESEM images to extract the thicknesses of LbL-assembled MXene films and extrapolated a thickness growth of around 9 nm per bi-layer (Table S1).[17] The linear growth, as a function of n, allowed us to use LbL-assembly for the fabrication of thin films with variable, precisely controlled thicknesses.

### 2.2 Spectroelectrochemical analysis

We performed in-situ UV-vis-NIR spectroelectrochemistry to characterize the electrochromic properties of the MXene films either spin-coated or LbL assembled on indium tin oxide (ITO) coated glass using a 100 mg mL$^{-1}$ poly(vinyl alcohol) (PVA)/ 6 v/v% $H_2SO_4$ gel electrolyte in water (Figure 2a, see the experimental section for details). We used an Ag/AgCl pellet as a pseudo-reference electrode to match the electrochemical potential between the spectroelectrochemical and the electrochemical transistor measurements (described in section 2.3), while a platinum coil was used as the counter electrode.

UV-vis-NIR absorption spectra of the pristine $Ti_3C_2T_x$ MXene thin films exhibit several characteristic features depending on the fabrication method.

The UV-Vis-NIR spectra of spin-coated MXene films exhibit a broad absorption spectrum, with a continuous increase of absorbance from the near-IR region to the UV attributed to the inter-band transitions[20] and a shoulder at around 760 nm (Figure 2b). Decreasing the number of spin-coating steps leads to a minor increase in the absorption at 760 nm with respect to the local maximum at λ < 450 nm (Figure S1a). In contrast to the spin-coated films, LbL-assembled MXenes with 3, 10, and 15 bilayers (Figure 2c, 2d, and S1b) exhibit a distinct band with an absorbance maximum at 760 nm attributed to surface plasmons, i.e., the collective oscillations of free electronic charge carriers.[21] Such a maximum is red-shifted to 807 nm for 20 bilayers (Figure S1c), indicating changes in MXene organization upon the increase in the number of layers. Overall, the increase in the number of LbL-assembled bilayers from 3 to 20 MXene bilayers leads to an enhancement in the magnitude of absorption of the films, consistent with an increase in thickness.

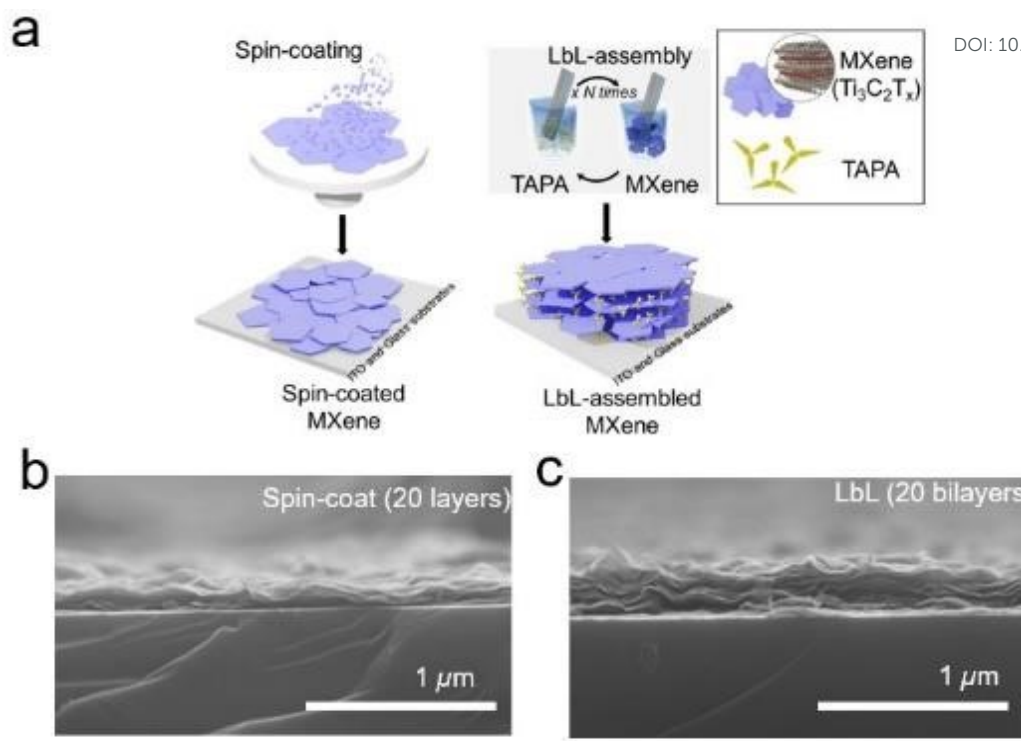

**Figure 1.** (a) Schematic diagram showing fabrication of MXene films using spin-coating, or using LbL self-assembly. SEM images of Mxene films fabricated with (b) 20 layers of spin-coated MXene and (c) 20 bilayers of MXene/TAPA LbL-assembled.

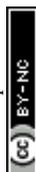

Application of cathodic biases from 0 V to -0.8 V leads to very small changes in the absorption spectra of the spin-coated films, with an overall decrease in absorption over the whole spectral range.

In contrast, layer-by-layer films, held together by TAPA in between the layers, exhibited a much larger change in absorption spectra for the same potential range. For 3, 10, and 15 LbL-assembled films, the local maximum below 500 nm exhibits a decrease in intensity, while the maximum at 760 nm showed a decrease in intensity and a blue shift to 670 nm. A similar trend was observed for 20 bilayers (Figure S1c). This phenomenon has previously been attributed to a change in the electron density of MXene, which in turn leads to a change in surface plasmon resonance.[22] Moreover, we observed an increase in absorption at wavelengths above 900 nm with the formation of an isosbestic point, attributed to the protonation of Ti=O functionalities on MXene surface to Ti-OH in the presence of strong acids, changing the oxidation state of Ti, and electron density.[22-24] Similar effects were previously observed for (MXene/TAPA).[19]

Figure S1d shows the relative changes in the absorption maximum of the plasmonic band for four different LbL-assembled films. The data show that, for up to 15 bilayers, there is an increase in the absorbance shift in the plasmonic band between the reduced (-0.8 V) and neutral state (0 V). Increasing further the number of bilayers to 20 resulted in a further decrease in absorbance shift, indicating that there is a threshold after which the electrochromic properties of the film are not maintained.

Spectroelectrochemical data at anodic potentials up to 0.6 V show an irreversible decrease in absorption over the whole spectral window, which may be attributed to the irreversible oxidation of $Ti_3C_2T_x$ (Figure S2).[22, 25] Larger changes could be observed for the films deposited via layer-by-layer assembly with respect to the spin-coated films, which could be ascribed to a disruption of the electrostatic interaction between the layers upon intercalation of sulfonate groups at oxidative potentials.[26]

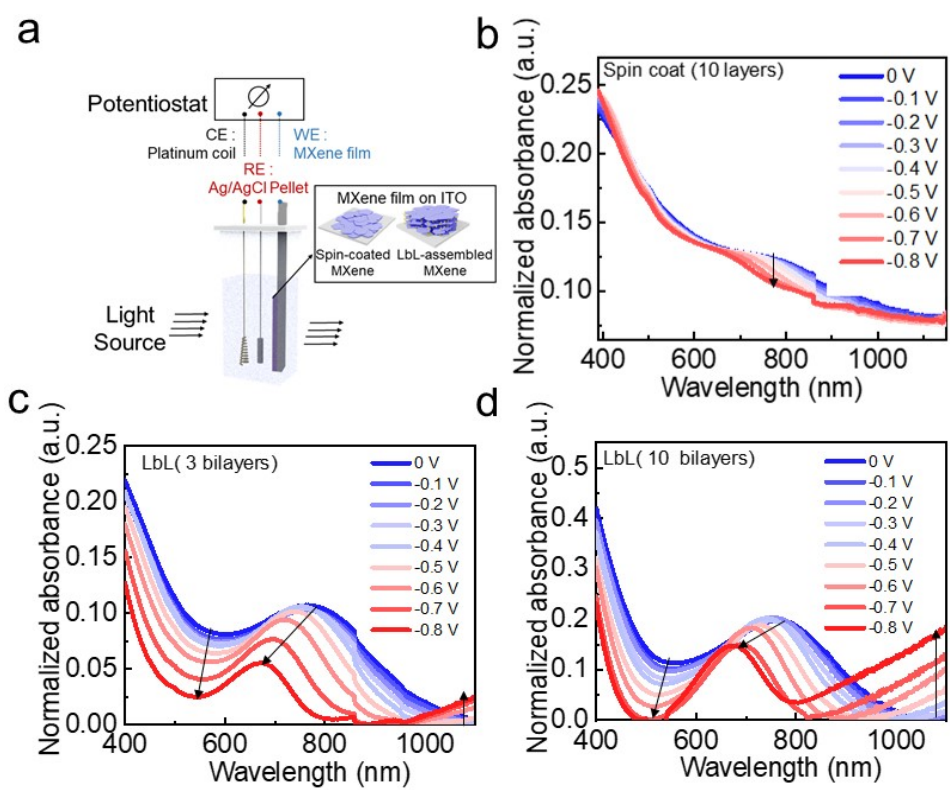

**Figure 2.** (a) Schematic diagram of the in-situ spectroelectrochemical measurement setup. UV-vis spectroelectrochemical measurements of (b) spin-coated (10 layers) and LbL-assembled (c) 3 bilayers (d) 10 bilayers MXene films at decreasing redox doping voltages from 0 to -0.8 V with steps of 0.1 V.

### 2.3 MXene ECTs

Contrary to spin-coated pure MXene films, LbL-assembly of MXene/TAPA layers allowed controlled growth and resulted in structured films showing larger changes in UV-Vis absorption spectra for potentials in the range of 0 to -0.8 V than spin-coated MXene films. Thus, we focused our attention on fabricating electrochemical transistor (ECT) devices using these LbL films. ECTs were fabricated using gold as electrode contacts with channel dimensions of 20 µm length and 1000 µm width. LbL-assembled films were used as the channel material. We used 100 mg mL$^{-1}$ PVA/ 6 v/v% $H_2SO_4$ as the gel electrolyte and an Ag/AgCl pellet as the gate (Figure 3a, see experimental section for details).

MXene/TAPA ECTs operate at positive drain voltages, consistent with the n-type transport of MXene (Figure S3). When the drain voltage is swept from 0 to 0.5 V, we observed that positive gate voltages lead to an increase in device current, consistent with increase of device current upon penetration of positive proton ions (Figure S4a). Application of negative gate voltages lead to a complete de-doping of the channel (at -0.35 V, Figure S4b), consistent with sulfonate ions penetration within the bulk of the film. However, this leads to a complete loss of device current, likely due to irreversible changes in the microstructure of the film upon oxidation, consistent with spectroelectrochemistry data (Figure S2). Similar phenomena can be observed when the device is operated at negative drain voltages (Figure S4c and S4d). In this case, higher gate voltages need to be applied in order to observe similar changes in drain current, consistent with the n-type character of MXene ECTs.[13]

The output characteristics show a linear increase of drain current with the drain voltage ($V_D$), with an ohmic behavior. Although The application of gate voltages from 0 to 1 V results in a further increase in channel current, consistent with spectroelectrochemical measurements and the protonation of the MXene layer upon application of positive gate voltages. Data on batch-to-batch device variability are shown in Figure S5 and Table S1.

Transfer characteristics for ECTs at drain voltage $V_D = 0.5$ V with varying channel thickness based on the number of MXene/TAPA bilayers are presented in Figures 3b-3d and S6. For all ECTs, gate current (Figure S7) values are in the range of 1-10 μA, at least one order of magnitude lower with respect to the drain current values, indicating that device operation is indeed dominated by the electronic conductivity of MXene (Figure S8). Transfer characteristics of all devices show hysteresis between the forward and reverse sweep (Figure S8), which could be due to irreversible processes or trapping of electrolyte ions within the polymer film.[27]

MXene/TAPA ECTs with 1, 3, and 5 bilayers show current modulation at gate voltages between 0 to 1 V and switch ON at $V_{TH} = 0.2$ V (determined using extrapolation in the linear region method). The sub-threshold regime for 5 bilayers MXene ECTs is shown in Figure S9. A larger number of bilayers resulted in an increase in threshold voltage (Table S1).[28] In contrast, the gate voltages needed for device switching and the extent of current modulation differ considerably for spin-coated devices (Figure S10) which turn ON at $V_{TH} = -0.3$ V and show current modulation at gate voltages between -0.3 to 0.3 V.

Output and transfer characteristics show that increasing the number of MXene/TAPA bilayers from 1 to 12 leads to an overall increase in device current at zero gate bias. We attribute the higher OFF current to the increase in film thickness, and thereby bulk electronic conductivity, with the number of bilayers.

Another parameter that is used to characterize ECTs is the transconductance ($g_m$), defined as the derivative of the drain-source current with respect to the gate voltage $(\partial I_D / \partial V_G)$. The transconductance is a measure of how much the device can amplify small voltage changes at the gate electrode.

The maximum average transconductance $g_{m,max}$ was 0.07 mS for n=1 bilayer and increased to $g_{m,max}= 2.2$ mS for n=3. Increasing further the number of bilayers to 5 led to a $g_{m,max}= 2.8$ mS (Figure 3b-d). However, we observed large batch-to-batch variations for films with 5 bilayers (Table S1). Increasing further the number of bilayers (7, 10, and 12) resulted in a decrease in transconductance (Figure 3e, and S6). This can be attributed several factors, that we will elaborate more on in our discussion about mobility. Another feature observed in some of the transconductance plots is that $g_m$ exhibit two maxima instead of one. This feature could be due to a slow response of MXene ECTs, leading to a non-monotonic change in drain current upon increasing the drain voltage. To ensure that this is the case, we extracted the ON and OFF time constant for 3 and 10 bilayer ECTs, exhibiting two and one $g_{m,max}$, respectively. The data confirm that for 3 bilayers $\tau_{ON} = 1.29 \pm 0.45$ s and $\tau_{OFF} = 0.40 \pm 0.05$ s, while for 10 bilayers $\tau_{ON} = 0.45 \pm 0.02$ s and $\tau_{OFF} = 0.42 \pm 0.03$ s (Figure S11), indicating that indeed the doping process is slower for thinner films. While this effect is not common with standard organic ECTs, where slower responses are generally observed for thicker films, previous work on layer-by-layer assembled polyelectrolyte multilayers showed that the surface redox potential varies as function of the polyelectrolyte layer number.[29] Such changes are expected to impact $H^+$ penetration within the bulk of the LbL film and, consequently, ECTs response times.[30]

Another parameter that could limit device performance is the type of material used as the gate electrode. To assess the impact of the gate on MXene ECT performance, we evaluated the effect of a true Ag/AgCl reference electrode (3 M KCl aqueous electrolyte) and free-standing MXene-CNF membrane as the gate materials (Figure S12 and experimental details section 4.9). Data show that the Ag/AgCl reference provides lower modulation with respect to the Ag/AgCl pellet, with $g_{m,max} = 0.8$ mS, indicating that the gate has a negative impact on the steady-state response of MXene ECTs, likely due to the mismatch in electrolyte. MXene-CNF gate provided smaller hysteresis than the Ag/AgCl pellet, and intermediate $g_{m,max} = 2.3$ mS, indicating a lower extent of channel doping with respect to the Ag/AgCl pellet, but a better reversibility of the doping process. Overall, the data show that the gate has a strong impact on the performance of MXene ECTs and that gates with large capacitance are to be sought to fully dope/de-dope MXene channels.

A key figure of merit used to benchmark materials with mixed ionic/electronic conductivity is μC*, where μ is the mobility and C* is the volumetric capacitance. Figure 4a shows that the measured areal capacitance is a linear function of thickness, with the slope corresponding to the volumetric capacitance C*. The volumetric capacitance C* extracted from cyclic voltammetry (CV) measurements at different LbL

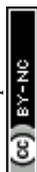

film thicknesses is 1364 F cm$^{-3}$ (Figure 4a and Figure S13),[31-33]

Although organic ECTs are different from FETs, it has been shown that the same equation as those used to describe electronic charge transport in long-channel FETs can be used for organic ECTs in steady-state.[34-36]. Specifically, the equation that describes transfer characteristics in a regime where doping occurs everywhere in the channel[36] is given by:

$$g_m = \frac{Wd}{L}\mu C^*(V_G - V_{TH}) \qquad (1)$$

where $g_m$ is the transconductance, $V_{TH}$ is the threshold voltage, $V_G$ is the gate voltage, $C^*$ is volumetric capacitance, and $\mu$ is the charge carrier mobility (electrons or holes).

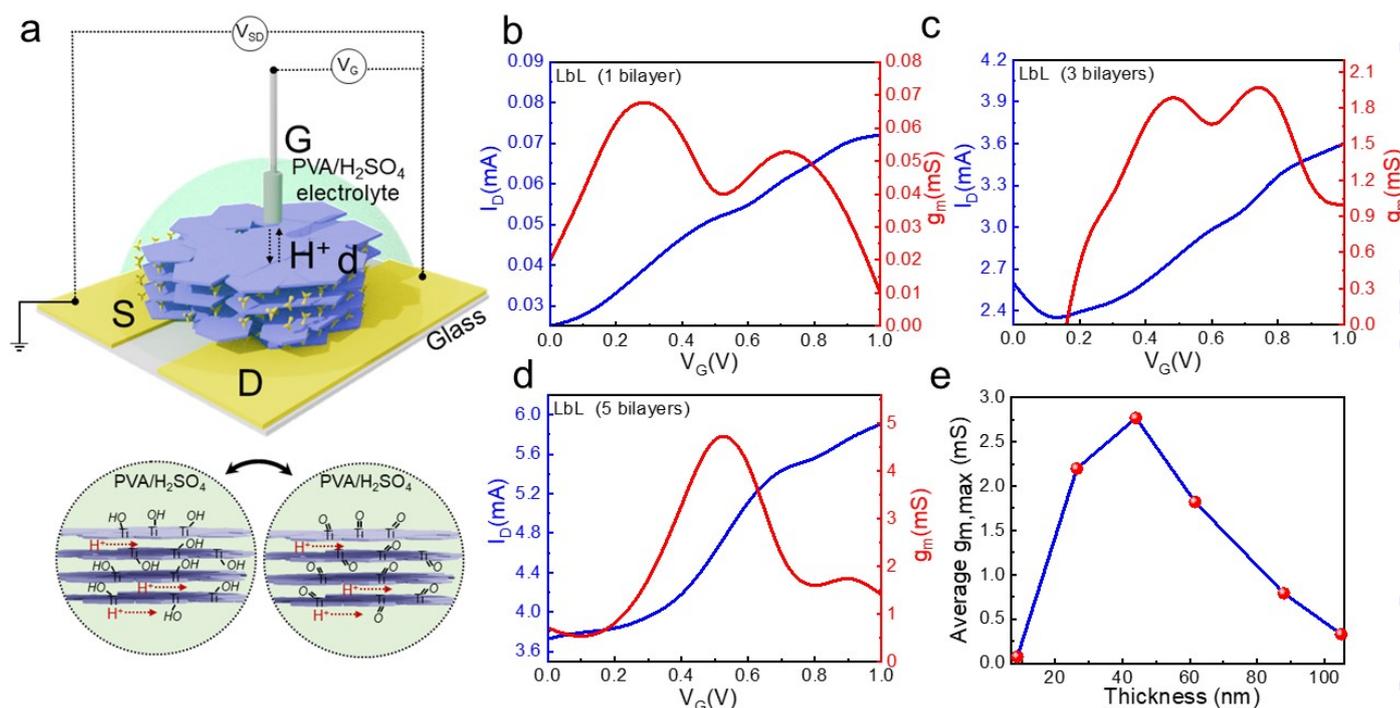

**Figure 3.** (a) Schematic representation of the ECT device. The transfer curves and calculated transconductance value of MXene ECTs having (b) 1 bilayer, (c) 3 bilayers (d) 5 bilayers as channels, measured for gate voltages ranging from 0 to 1 V and $V_D$ = 0.5 V. (e) The variation of average maximum transconductance with respect to the channel thickness.

The main difference between the equation for ECTs and FETs is that the transconductance in ECTs scales with the thickness and not just the channel area (i.e., *W* and *L*). This distinction arises since field-effect only modulates the channel at the semiconductor-insulator interface whereas redox doping modulates the carrier density throughout the bulk of the ECT channel.[35, 36]

We therefore used this equation to calculate the mobility µ of the different films in the transistor (Figure 4b). In addition to µ, Figure 4b also displays the variation of maximum channel current with respect to *d*, showing a linear increase with thickness. The maximum calculated mobility was 0.036 cm$^2$ V$^{-1}$ s$^{-1}$ – a comparable value to that measured for state-of-the-art organic ECT materials.[37,38] The mobility values, however, decreased with a further increase in thickness (above 5 bilayers). Since the capacitance, and maximum channel current both show a linear increase with thickness, the decrease in mobility can only be related to the decrease in transconductance. Data show that indeed $g_m$ starts decreasing for thicknesses above 45 nm (Figure 3e). Previous reports have shown that parasitic contact resistance limits $I_{D,max}$, and consequently $g_{m,max}$,[37] for large $W_d/L$ values. For the 2D ECTs reported here, W/L was 1000/20 µm, which could lead to saturation even at relatively low thicknesses above 45 nm. To verify this hypothesis, we have tested MXene ECTs (5 bilayer) with different W/L ratio (Figure S14). The data show that the maximum drain current and device hysteresis increase with the W/L ratio. Most importantly, the data confirm that $g_m$ saturates for high W/L ratio, corroborating our hypothesis that contact resistance is the main factor behind $g_m$ saturation at high $W_d/L$ values.

We note however that the mobility values are already good enough for some applications. The maximum ECT transconductance is also quite high, with an average value of 2.2 mS for 3 LbL MXene ECTs. Since MXene ECTs also operate at low threshold voltages $V_{th}$ and the redox state of the channels is stable even when the gate voltage is turned off, these devices have low energy consumption when used as low power memory components, as in Electrochemical Random-Access Memories (ECRAM) for neuromorphic devices.

Stable operation is an important parameter to consider when assessing device performance. We assessed the operational stability of 5 bilayers MXene ECTs by applying a square wave voltage to the gate electrode (from $V_G = 0$ to 0.6 V) while monitoring the drain current at a constant drain voltage of $V_D = 0.5$ V (Figure S15). MXene ECT devices show around 94.4% retain in maximum drain current after 100 cycles.

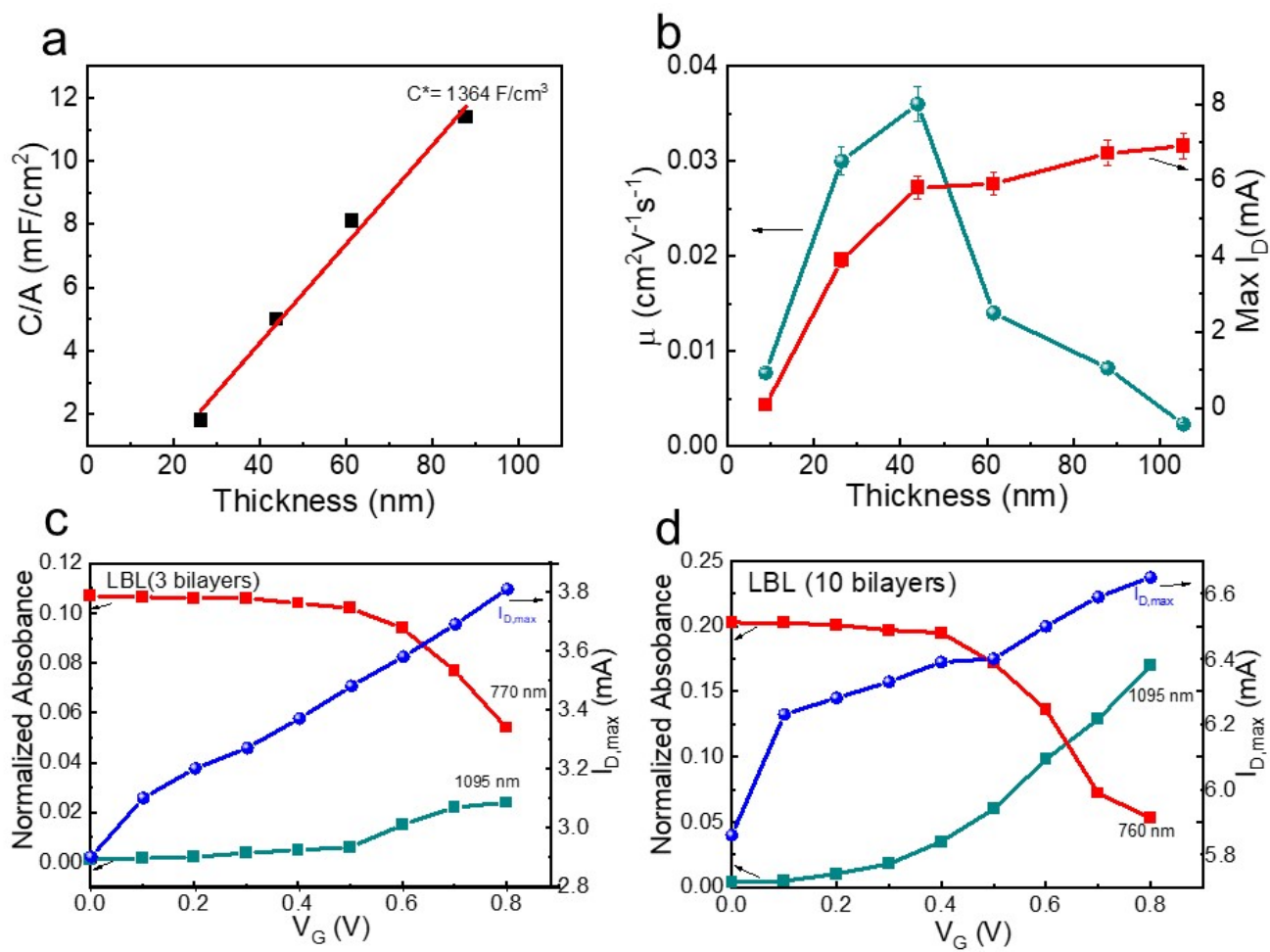

**Figure 4.** (a) Volumetric capacitance C* extrapolated from capacitance versus thickness plot for LbL films with 3, 5, 7 and 10 bilayers, measured for channel geometry of L = 20 µm and W = 1000 µm. Capacitance values are extracted from cyclic voltammetry data using Equation 2. (b) Calculated mobility and measured maximum current for LbL assembled MXene with varying thickness. (c-d) ECT Drain current vs. UV-vis absorbance at 1095 nm and 770 nm measured at different redox potentials (vs. Ag/AgCl) for film of different LbL bi-layer number.

## ARTICLE

### 2.4 The origin of conductance switching in MXene: Correlating channel conductivity to electrochromism

Many mixed ionic/electronic conductors exhibit electrochromism,[39] a reversible change in their optical properties due to a change in their redox state.

Previous work has shown that $Ti_3C_2T_x$ MXene thin films can undergo reversible changes in optical absorption upon electrochemical cycling in acidic electrolytes ($H_3PO_4$ and $H_2SO_4$) at potentials below 1 V (with silver wire as the reference electrode).[40, 41] Some MXenes may exhibit metal-insulator transition meditated by adsorbates induced by electrochemistry on their surface.[42]. Such electrochromic effect is attributed to a non-faradaic electrical double layer as well as a faradaic redox process.[43] In the case of acidic electrolytes, the main effect should be from the proton redox of Ti. To assess whether MXene ECTs could also operate in contact with aqueous electrolytes, we acquired 5 bilayers MXene ECTs using 0.1 M $NaCl_{(aq)}$ as the electrolyte (Figure S16). Data show that the aqueous electrolyte leads to similar MXene ECT operation with respect to the acid electrolyte, leading to an increase in drain current upon application of positive gate voltages. However, these devices led to maximum transconductance $g_{m,max}$ of 0.93 mS and $I_{on}/I_{off}$ ratio of 1.4, with respect to $g_{m,max}$ of 2.76 mS and $I_{on}/I_{off}$ ratio of 1.58 observed for PVA-$H_2SO_4$ electrolyte. Overall, the data indicate that $H^+$ are more effective than $Na^+$ as doping ions in MXene ECTs.

For organic ECTs, it is known that ion-electron coupling, i.e. redox reactions, leads to the formation of charge carriers (e.g., polarons and bipolarons) in the conjugated polymer backbone, resulting in a change in its bandgap, and its conductivity.[6] Here, we further correlate the phenomenon of changes in film conductance as observed in ECTs, by linking it to the optoelectrochemical behavior of LbL-assembled MXene.

To further consolidate our results, we compared the changes in optical absorbance at selected wavelengths as obtained from spectroelectrochemical measurements to the drain current obtained from ECT transfer characteristics (Figure 4c, 4d). Previous work showed that redox potentials extracted from in-situ UV-Vis spectroelectrochemistry correlate well with threshold voltage values extracted from ECT measurements.[44, 45, 46] To ensure that the electrochemical potential can be matched between the experiments we used the same electrolyte and Ag/AgCl pseudoreference electrode for both types of measurements (see schematic figure 2a, and 3a). The data show that the increase in drain current coincides with a decrease in absorbance of the plasmonic band (760 nm) and an increase of absorbance at wavelength 1095 nm. These data suggest that the reversible protonation of the Ti (see schematic Figure 3a) is both responsible for electrochromism as well as conductance change upon redox doping of the MXene film. This is to our knowledge the first report of correlating optoelectronic properties and electrochemical device performance in MXene films.

In addition, during ion intercalation processes, ions could physically expand the interlayer spacing, which may also lead to a change in the property of the films. All these phenomena combined lead to changes in the overall optical (as probed by UV-Vis) as well as long-range electronic conductivity of the ECTs. Stability data show that this is indeed the case for LbL-assembled MXene ECTs, leading to a decrease in stability upon ON/OFF switching at $V_G$ from 0 to 0.6 V (Figure S15). Future work should focus on a deeper understanding of these phenomena, which could be further modulated by the type of molecule used as an interlayer between MXene flakes as well as the chemical structure of MXene.

### 3. Conclusion

The solid-state field-effect transistor, FET, and its theories were paramount in the discovery and studies of graphene. Here we show that MXenes, a class of 2D materials beyond graphene, have mixed electronic ion properties that enable the realization of electrochemical transistors (ECTs). The current parameters of the MXene ECTs show good normalized transconductance (12.5 S.cm$^{-1}$) but low on-off ratios (1.19), these values are already state of the art for use in devices like electrochemical Random-Access Memories (ECRAMs) or biosensors.

We further show that conductance switching data measured using ECTs in combination with other in-situ electrochemical measurements (such as UV-Vis), is a powerful tool to correlate the changes in device current to that of redox states: to our knowledge, this is the first report of this important correlation for MXene films.

Further studies beyond graphene 2D materials within the framework of ECTs are expected to enable the optimization of parameters, such as higher on-off ratios, very large channel currents, increasing transconductance, or extremely high switching speeds. Future experiments on Donnan permselectivity of LbL MXene with different types of cationic spacers are expected to elucidate the impact of redox formal potential in LbL films on ECT operation.[29] Spatiotemporal in-situ optical measurements of the channel area will provide information on whether the so-called moving front effect,[44] which has been shown for organic electronic material, is also present for 2D MXene or other 2D materials.

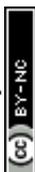

Many future possibilities exist for MXenes ECTs as over 30 different MXenes are predicted. We thereby foresee that even transition metal dichalcogenides (TMDs) and other bandgap materials can form ECTs and 2D iontronics. Future ECTs may also comprise heterostructure films with mixture of MXenes and 2D TMDs.

2D ECTs can draw great inspiration and theoretical tools from the field of organic ECTs.[47-49] They have the potential to considerably extend the capabilities of transistors beyond that of organic ECTs, as they have added properties such as extreme heat resistance, tolerance for solvents, and higher conductivity for both electrons and ions than organic mixed ionic/electronic conductors, such as conducting polymers.

## 4. Methods

### 4.1 Synthesis of MXene

MAX phase was synthesized according to our previous work[17] and powdered. MAX phase ($Ti_3AlC_2$) powder was delaminated in 2 g of LiF completely dissolved in 40 mL of 6 M HCl solution under constant stirring for 15 mins. This was followed by the addition of 2 g of MAX phase powder at the rate of 200 mg/min in an ice bath. The etchant was maintained at 50 °C for 24 h at 550 rpm under constant stirring. The etchant was cooled to room temperature and washed several times with Millipore water until it reached pH≈6. The obtained powders were dispersed in deaerated Millipore water and probe sonicated for 1 h in an ice bath. The dispersed solution was centrifuged at 3500 rpm for 1 h, the supernatant collected, centrifuged at 5000 rpm, and re-dispersed in 10 mL Millipore water using vigorous shaking for 15 minutes to form the final MXene ink. The ink concentration was determined from the net weight of MXene film obtained by vacuum filtration of a known volume of ink on a Celgard membrane.

### 4.2 MXene characterization using SEM

We produced multilayered MXene films via both spin-coated and LbL-assembled technique on Si substrates for SEM observations. Cross-sectional SEM images were obtained by field emission SEM (Hitachi S4800, Hitachi Corp., Japan).

### 4.3 PVA-H2SO4 Gel Electrolyte Synthesis

To purify absorbed air in the water, we bubbled 10 mL Milli-Q water with Ar gas for 1 h and used it to dissolve 1 g (polyvinyl alcohol) PVA (Sigma, Mw = 89 000–98 000, 99+% hydrolyzed). The solution was stirred at 85 °C on a hot plate for at least 3 hours until it became transparent, indicating that PVA is well dissolved. Sequentially, the prepared solution was removed from the hot plate and cooled down to room temperature. Concentrated $H_2SO_4$ (0.6 ml) (Sigma, >97.5%) was very slowly added to the PVA solution, followed by stirring for at least 1 h at room temperature.

### 4.4 ECTs fabrication

We utilized two types of MXene multilayers as redox-active channels in our study: LbL-assembled and spin-coated MXene multilayers. Prior to multilayers formations, bare glass substrates were bath-sonicated in the sequence of acetone, ethanol, and isopropyl alcohol for 20 min each. Afterward, we patterned them with Cr (50 nm) and Au (150 nm) electrodes (channel dimensions, L = 20 µm, W = 1000 µm) using photolithography. Then, we treated $O_2$ plasma (Optrel GBR, Multi-stop) for 20 min to create the hydrophilic surface to facilitate the uniform coverage of 2D MXene for both cases.

To avoid any parasitic capacitances from metal/electrolyte interfaces, we coated polymethyl methacrylate (PMMA) on the source-drain Au electrode after LbL-assembled MXene formation to complete ECT devices. To prepare PMMA solution, we dissolved 1 g poly(methyl methacrylate) (Sigma, Mw ≈ 996 000 by GPC, crystalline) in 10 mL Toluene (Sigma, ACS reagent, ≥99.5%), and the mixed solution was stirred vigorously at 80-degree Celsius overnight.

For the spin-coated MXene multilayers, aqueous $Ti_3C_2T_x$ MXene solution (2 g L$^{-1}$) was coated at 3000 rpm for 30 seconds. This step was repeated 5, 10, or 20 times to form multilayers of increasing thickness.

In the case of LbL-assembled MXene multilayers, we used tris(3-aminopropyl)amine (TAPA, Sigma) as a spacer molecule to provide the efficient interlayer chemistry. We prepared 1 bilayer (8.79 nm), 3 bilayers (26.38 nm), 5 bilayers (43.97 nm), 7 bilayers (61.55 nm), 10 bilayers (87.94 nm) and 12 bilayers (105.5 nm) of MXene-TAPA bilayer. Prior to dipping the substrates, we used 3M adhesive tape to prohibit the back-side coating of the substrate during dipping process. We treated the substrate surface by using $O_2$ plasma for 15 mins to modify its wetting properties and as-treated substrates were then loaded into a dipping robot (StratoSequence VI, nanoStrata Inc.). For LbL assembly setup, aqueous $Ti_3C_2T_x$ MXene dispersion (2 g L$^{-1}$) and TAPA dissolved in milli-Q water (1 g L$^{-1}$) were used. We did 8 consecutive dipping steps including rinsing with pure milli-Q water using an automated program (StratoSmart v7.0). For the one cycle, i.e., one bi-layer formation, the substrates were firstly dipped into TAPA-dissolved solution and sequentially spun in a circle for 5 min. Subsequently, they were rinsed with the following 3 steps for 2 min each. After the formation of positive charges on the surface, the substrates were dipped and spun again in the negatively charged MXene solution for 5 min and rinsed repeatedly 3 times for 2 min. These steps were repeated for the formation of MXene multilayers. We then dried the LbL assembled films overnight in a vacuum oven at room temperature. To prevent any parasitic capacitances from undesired Au bottom electrode, we drop-casted a solution of PMMA, Sigma, Mw ~996,000 by GPC, crystalline) on the edge of dried MXene film. The gel electrolyte was then drop-casted on top of LbL-assembled MXene multilayers. Finally, Ag/AgCl pellet was dipped in the gel electrolyte to complete the electrochemical transistor configurations.

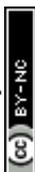

### 4.5 ECT data collection and characterization

We measured transistor data using a Keithley 4200A-SCS parameter analyser in the air at room temperature. Data analysis was performed using Origin Pro 2020. Transconductance data (calculated by using the equation $g_m = \partial I_D / \partial V_G$) were smoothed by adjacent averaging to attenuate instrument and environment noise. We calculated the threshold voltage using Extrapolation in the linear region method (ELR) method. This method calculates the threshold voltage by finding the gate voltage axis intercept of the linear extrapolation of the $I_D$ versus $V_G$ curve at its maximum first derivative point.

### 4.6 In-situ spectroelectrochemistry

An ITO glass slide coated with MXene was served as the working electrode. Pt coil was used as the counter electrode and Ag/AgCl pellet was used as the reference electrode. The three-electrode cells were dipped together in the 1 M PVA-$H_2SO_4$ electrolytes in a standard quartz cuvette, located in the middle of beam path of an ultraviolet–visible (UV–vis) spectrophotometer (PerkinElmer LAMBDA 750 UV/Vis/NIR Spectrophotometer). For in-situ setup, we connected three electrodes with cyclic voltammetry (CV) and recorded the corresponding optical absorption as a function of applied potential with a step of 0.1 V.

### 4.7 Impedance measurements

MXene films were deposited via LbL assembly on square gold electrodes of dimension 2 mm$^2$. For electrode fabrication, a glass wafer was cleaned via subsequent ultrasonication in soapy water for 15 minutes, basic piranha cleaning in a solution of $H_2O$:$H_2O_2$:$NH_3$ at around 70 ºC for 5-10 minutes, and $O_2$ plasma for 15 mins. Gold contacts were prepared by evaporation using a custom-made paper mask[50] to form 10 nm of chromium as an adhesion layer and 50 nm of gold.

EIS measurements were performed on LbL-assembled MXene using a BioLogic VSP potentiostat. The MXene films on gold were used as the working electrode, an Ag/AgCl pellet pseudo reference was used as the reference electrode, and a platinum coil as a counter electrode The electrolyte was 1 M poly(vinyl alcohol) (PVA) in $H_2SO_4$, as for the ECTs. EIS data were acquired at frequencies between 100 kHz to 10 Hz at a sinus amplitude of 10 mV and a potential of -0.8 V. The capacitance was extracted by fitting the data to a modified Randles circuit model Rs(Rp||Q), where Rs is the electrolyte resistance, Rp is the charge transfer resistance, and Q is a constant phase element, using ZFit (EC-Lab V11.41 software). The thickness of the films was determined in a dry state from the SEM cross section images.

### 4.8 CV measurements and calculation of capacitance

CV measurements were performed on LbL-assembled MXene using a BioLogic VSP potentiostat with the same setup used for EIS measurements. The CV curves were obtained at a scan rate of 100 mV/s. The specific capacitance C (Fcm$^{-2}$) was estimated by using equation 2[51]

$$c = \frac{1}{2kA\Delta V} \int I dv \qquad (2)$$

where k is the scan rate (V s$^{-1}$), A is the Area of the active material (cm$^2$), $\Delta V$ is the potential window (V) and IdV represents the area under CV curve.

### 4.9 Free standing MXene membrane

MXene membranes were prepared by mixing MXene with cellulose nanofibers (CNF). The MXene-CNF dispersions were vacuum-filtered into sheets using a microfiltration assembly with Durapore PVDF membranes and dried for 15 min using a Rapid Köthen sheet drier (Paper Testing Instruments, Austria) at 93 °C and a reduced pressure of 95 kPa. Dried sheets were cut into suitable sizes (length 2.5 cm, width 0. 5 cm) to use as gate for ECT devices.

## Conflicts of interest

There are no conflicts to declare.

## Acknowledgements

MinA acknowledges funding from ÅForsk project 18-461. J.S. acknowledges funding from Olle Engkvists stiftelse. M.H. acknowledges funding from energimyndigheten project 48489-1. E.Z. gratefully acknowledges the European Union's Horizon 2020 research and innovation program under the Marie Sklodowska-Curie grant agreement "BioResORGEL" (Program No. 101025599).

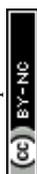

View Article Online
DOI: 10.1039/D3NR06540E

# ARTICLE